\documentclass{aa501}
\usepackage{times}
\usepackage{epsfig}
\begin{document}
%\thesaurus{(13.07.1; 02.18.5; 02.16.1; 09.10.1; 11.09.4; 11.19.3)}
\title{Colors and luminosities of the optical afterglows of the
$\gamma$-ray~bursts}
\author{V.~\v{S}imon \inst{1}, R.~Hudec \inst{1}, G.~Pizzichini \inst{2} 
\and N.~Masetti \inst{2} }
\offprints{V.~\v{S}imon: simon@asu.cas.cz}
\institute{Astronomical Institute, Academy of Sciences of the Czech 
        Republic, 251~65~Ond\v{r}ejov,
Czech Republic \and Istituto 
        Tecnologie e Studio delle Radiazioni
Extraterrestri, CNR, via 
        Gobetti 101, 40129 Bologna, Italy }
\date{Received date; accepted date}
\authorrunning{\v{S}imon et al.}
\titlerunning{Optical afterglows of GRBs}
%\maketitle
\abstract{
Results of  the  study  of the color indices and luminosities of 17 optical 
afterglows (OAs) of  GRBs  are presented. We show that the color variations 
during  the  decline  of  OAs  (except  for GRB000131) are relatively small 
during $t-T_{\rm 0}<10$  days and  allow a comparison  among them, even for 
the less densely  sampled OAs. The colors  in the observer frame, corrected 
for  the  Galactic   reddening,  concentrate  at  $(V-R)_0~=  0.40\pm0.13$, 
$(R-I)_0~=  0.46\pm0.18$,  $(B-V)_0~=  0.47\pm0.17$. The color evolution of 
the  OAs  is negligible  although  their  brightness  declines  by  several 
magnitudes during the considered time interval. Such a strong concentration 
of the  color indices  also  suggests  that the intrinsic reddening (inside 
their host galaxies) must be quite  similar  and  relatively  small for all 
these  events. The  absolute  brightness  of  OAs  in  the  observer frame, 
corrected for the host galaxy, lies within $M_{\rm R_0}~= -26.5$ to $-22.2$  
for  $(t-T_{\rm 0})_{\rm rest}=0.25$  days. This spread of $M_{\rm R_0}$ is  
not significantly  influenced  by  the  shifts  of $\lambda$, caused by the 
different  redshift  $z$ of the respective OAs. The general decline rate of 
the  OA  sample  considered  here  seems  to be independent of the absolute 
optical brightness of the OA, measured at some  $t-T_{\rm 0}$ identical for 
all OAs, and  the  light  curves  of  all events are  almost parallel, when 
corrected for the redshift-induced time dilation.
}

\maketitle

\section{Introduction}

     The color  indices  of  the optical afterglows (OAs) of the $\gamma$-ray 
bursts (GRB) are  a powerful  tool  to  use  in  the  search  for  the common 
properties  of  these  events. The OAs in  the  fireball  model represent the 
stadium when the matter from the central engine, moving  at  the relativistic 
speed, interacts  with  the  surrounding  interstellar  medium  by  means  of 
external  shocks (see  Piran (1999) for review). The color indices of the OAs 
can  be used  as an  important  parameter reflecting  the  related   physical 
processes. Besides the  astrophysical analysis, the specific color indices of 
OAs give hope to resolving whether an optical transient event is related to a 
GRB even  without available $\gamma$-ray detection (see also e.g. Hudec 2000, 
Rhoads 2001). The  redshift $z$, known for some OAs, can help assess the role 
of the  shifts  of  optical  passbands  with  respect  to the rest frame. The 
distance  of  OAs  can  be  determined from their redshift $z$ enabling their 
absolute brightnesses to  be calculated. The interrelations among the colors, 
luminosities  and  the  decay  rates  of  the  OAs  can thus be searched for. 
Moreover, we will show that  the colors of the OAs enable us to put stringent 
constraints on  the  properties  of the local interstellar medium of the GRBs 
considered here.

     The preliminary versions of this analysis were presented by \v{S}imon et 
al. (2000abc).

\begin{figure*}%[ht!]
%\rule{0.4pt}{13.1cm}
%\rule{0.4pt}{1cm}
\epsfig{figure=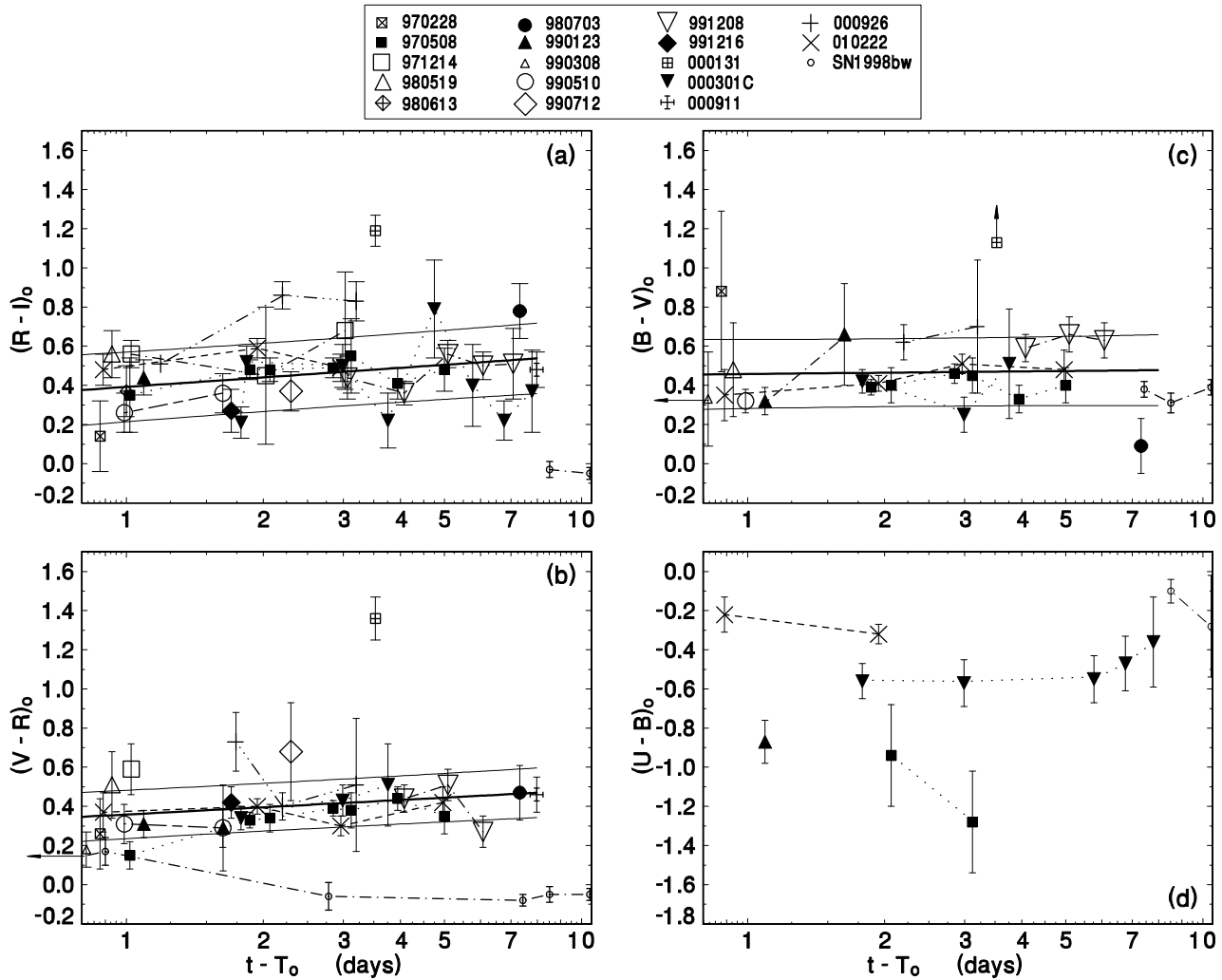,angle=0,width=18cm}
%\vspace{6.5cm}
\caption[ ]{Temporal  evolution  of the color indices of the respective 
afterglows. $t-T_{\rm  0}$ is the  time  interval in the observer frame 
elapsed  from  the  corresponding GRB. Colors  of all OAs correspond to 
their final decline branch  with  the exception  of  the first point of 
GRB970508. Only observations within $t-T_{\rm 0}<10$~days were plotted. 
The  error  bars  denote the standard deviation. The color indices of a 
given  OA  are connected by lines for convenience. The fit to the whole 
ensemble of the OAs, except for the largely outlying GRB000131, is also 
shown (thick solid line) along with its standard  deviation (thin solid 
line). The OA of GRB990308 ($t-T_{\rm 0}=0.14$~days) falls out of scale 
in Figs.\,bc but is included  in  the  fits. The colors of SN1998bw are 
shown just for comparison and are not included in the fits.  }
\label{t-color}
\end{figure*}

\section{Collection and analysis of the data   \label{col}  }

     This comprehensive analysis  has  made  use of the data published in 
the journals, the GCN  circulars\footnote{available   at  the  URL:\\{\tt 
http://gcn.gsfc.nasa.gov/gcn/gcn3\_archive.html}} and in J.~Greiner's Web 
page\footnote{{\tt  http://www.aip.de/People/JGreiner/grbgen.html}}.  The 
summary of  the  literature  of  the suitable optical data for the OAs is 
shown in Table~\ref{lit}. At present, suitable {\it multicolor} photometry 
is available  only  for 17 OAs. Moreover, the  published photometry often 
comprises  unorganized  observations. This fact seriously complicates the 
reconstruction  of  the  light  curves  in  the various passbands and the 
analysis of the color variations over the whole event.

\begin{table*}[t]
\caption[ ]{Summary of literature used for the construction of the light 
curves  of  the  optical  afterglows. Only  those GRBs, for whose OAs at 
least one color index for $t-T_{\rm 0}<10$ days could be determined, are 
listed. }   \label{lit}
\[
\begin{array}{ll}
\hline
\noalign{\smallskip}
$GRB970228$ & $Guarnieri et al. (1997), IAU Circ.6618, van Paradijs et al.
              (1997), Castander \& Lamb (1999), Pedichini$ \\ & $et al.
              (1997), IAU Circ.6588, IAU Circ.6631, IAU Circ.6619, Fruchter
              et al. (1999), Masetti et al. (1998)$           \\
$GRB970508$ & $Galama et al. (1998a), Zharikov et al. (1998), Sokolov et al.
              (1998)$                                         \\
$GRB971214$ & $Halpern et al. (1998), Diercks et al. (1998), Kulkarni et al.
             (1998), GCN 61, IAU Circ. 6793, Gorosabel$ \\ & $et al. (1998),
             Ramaprakash et  al. (1998)$   \\
$GRB980519$  & $Halpern et al. (1999), Vrba et al. (2000)$     \\
$GRB980613$  & $GCN 109, 117, 118, 134, 189$  \\
$GRB980703$  & $Bloom et al. (1998)$       \\
$GRB990123$  & $Galama et al. (1999)$      \\
$GRB990308$  & $Schaefer et al. (1999)$    \\ 
$GRB990510$  & $GCN 316, 318, 319, 321, 323, 324, 325, 328, 329, 330, 331,
               332, 386$ \\
$GRB990712$  & $GCN 389, 391, 402, 403, IAU Circ. 7225$        \\
$GRB991208$  & $Castro-Tirado et al. (2001), GCN 475, 481$     \\
$GRB991216$  & $Halpern et al. (2000), GCN 496$                \\
$GRB000131$  & $Andersen et al. (2000)$    \\
$GRB000301C$ & $Masetti et al. (2000a), Jensen et al. (2000)$  \\
$GRB000911$  & $Palazzi et al. (in preparation)$               \\
$GRB000926$  & $Price et al. (2001), GCN 807$                  \\
$GRB010222$  & $Masetti et al. (2001), GCN 1000, 1003, 1009$ \\
\noalign{\smallskip}
\hline
\end{array}
\]
\end{table*}

\begin{figure}
%\rule{0.4pt}{11.2cm}
%\rule{0.4pt}{1cm}
\hspace{-2cm}
\epsfig{figure=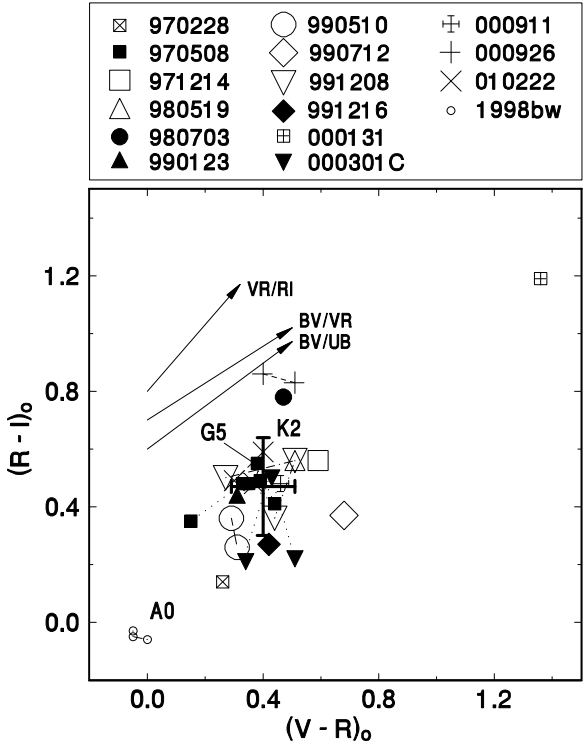,angle=0,width=13cm}
%\vspace{5cm}
\caption[ ]{$V-R$ vs. $R-I$ diagram of the afterglows. The color indices 
were  corrected  for the  Galactic reddening. Only  observations  within 
$t-T_{\rm  0}<10$~days were plotted. Multiple indices of the same OA are 
connected by lines  for  convenience. The  mean colors (centroid) of the 
whole  ensemble  of  OAs  (except  for  the largely outlying GRB000131), 
displayed in this diagram, including the standard deviations, are marked  
by the large cross. The colors of SN1998bw are shown just for comparison 
and were not included in calculation of the centroid. The representative  
reddening paths for $E_{\rm B-V}=0.5$ are also  shown. Positions  of the 
main-sequence stars are included for comparison. See text for details. } 
\label{vrri-aa}
\end{figure}

\begin{figure}
%\rule{0.4pt}{10.7cm}
%\rule{0.4pt}{1cm}
\hspace{-2cm}
\epsfig{figure=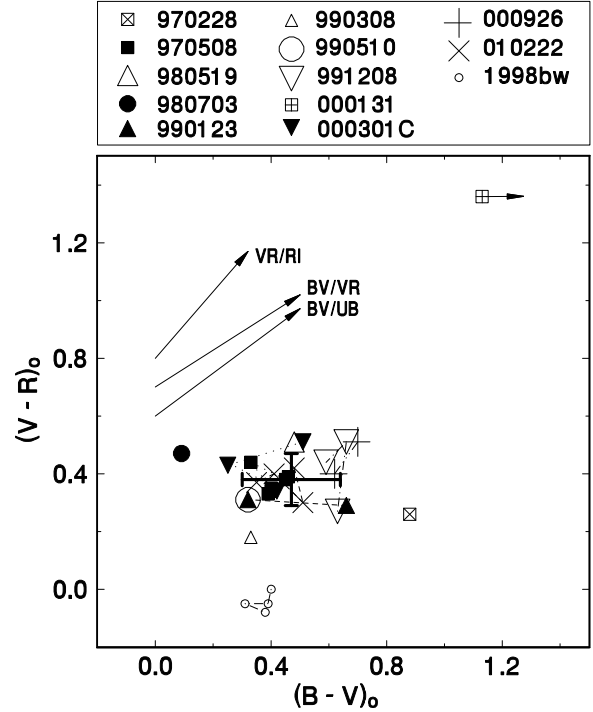,angle=0,width=13cm}
%\vspace{5cm}
\caption[ ]{$B-V$ vs. $V-R$ diagram of the afterglows. The arrangement 
is the  same  as  in Fig.\,\ref{vrri-aa}. The  colors of GRB000131 and 
SN1998bw are shown  just  for  comparison  and  were  not  included in 
calculation of the centroid.}
\label{bvvr-aa}
\end{figure}

\begin{figure}
%\rule{0.4pt}{8.5cm}
%\rule{0.4pt}{1cm}
\begin{center}
%\hspace{-3cm}
\epsfig{figure=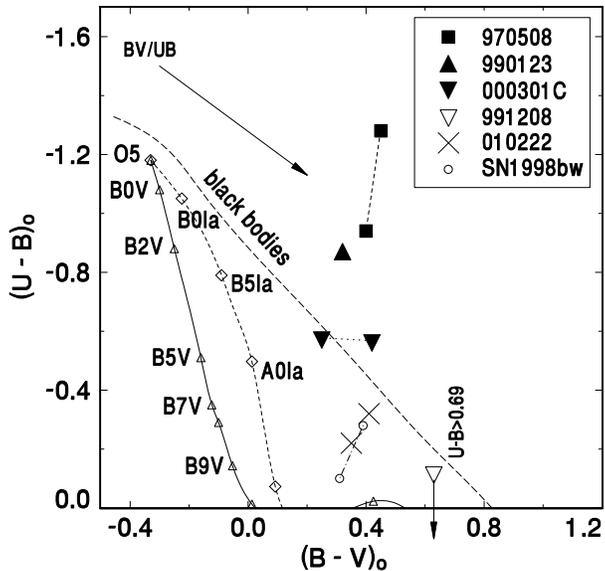,angle=0,width=10cm}
\end{center}
%\vspace{5cm}
\caption[ ]{$U-B$ vs. $B-V$  diagram  of  the  afterglows. Positions of 
the  main  sequence  stars, giants  and  the  locus of black bodies are 
plotted~-- they can be used for the {\it observational} differentiation 
of  OAs  from  other  kinds  of  objects. The arrangement is similar to 
Fig.\,\ref{vrri-aa}. See Sect.\,\ref{color} for details.}
\label{bvub-aa}
\end{figure}

\begin{figure}
%\rule{0.4pt}{16.7cm}
%\rule{0.4pt}{1cm}
\hspace{-2cm}
\epsfig{figure=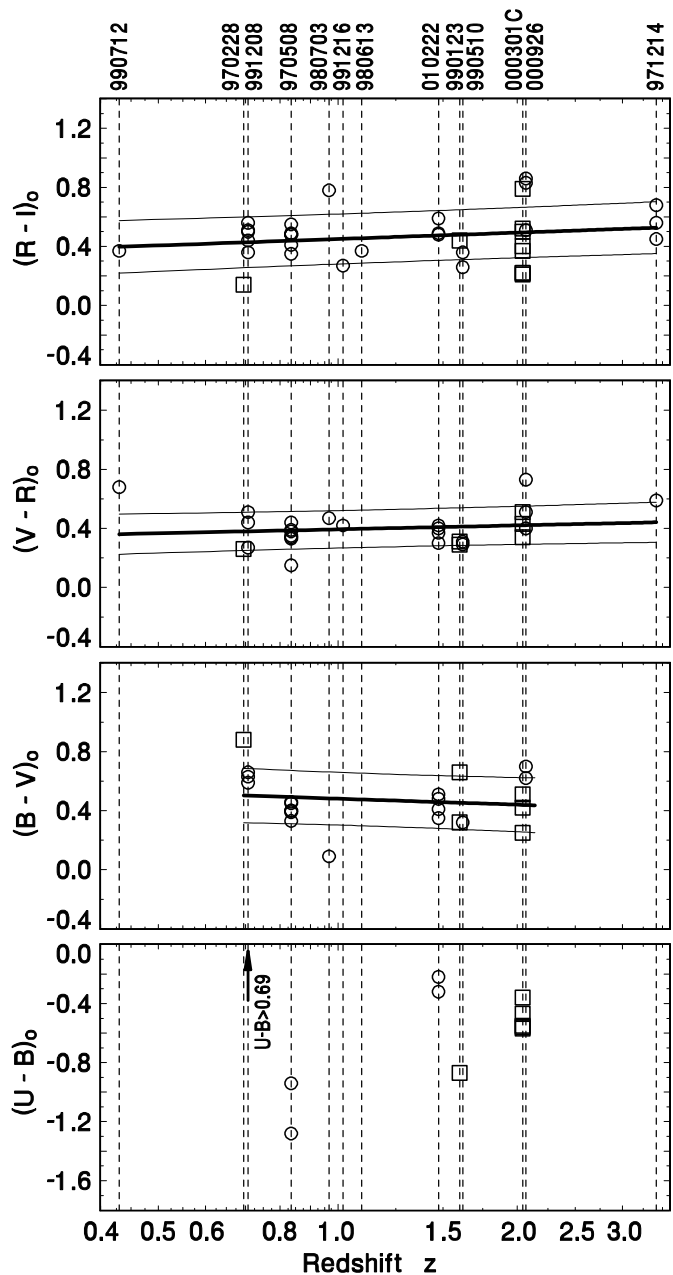,angle=-90,width=13cm}
%\vspace{-.5cm}
\caption[ ]{The color  indices  of all OAs with known redshift $z$ 
plotted  as a function  of  $z$. Only  the  OAs  with  $z<3.5$ are 
considered  here. The fits and their 1\,$\sigma$ errors are marked 
by  the  thick  and  thin solid lines, respectively. The weight of 
each  OA, used for  each  fit, was  proportional  to the number of 
available color indices. See Sect.\,\ref{red} for details.}
\label{color-z2}
\end{figure}

\begin{figure}
%\rule{0.4pt}{7.4cm}
%\rule{0.4pt}{1cm}
\epsfig{figure=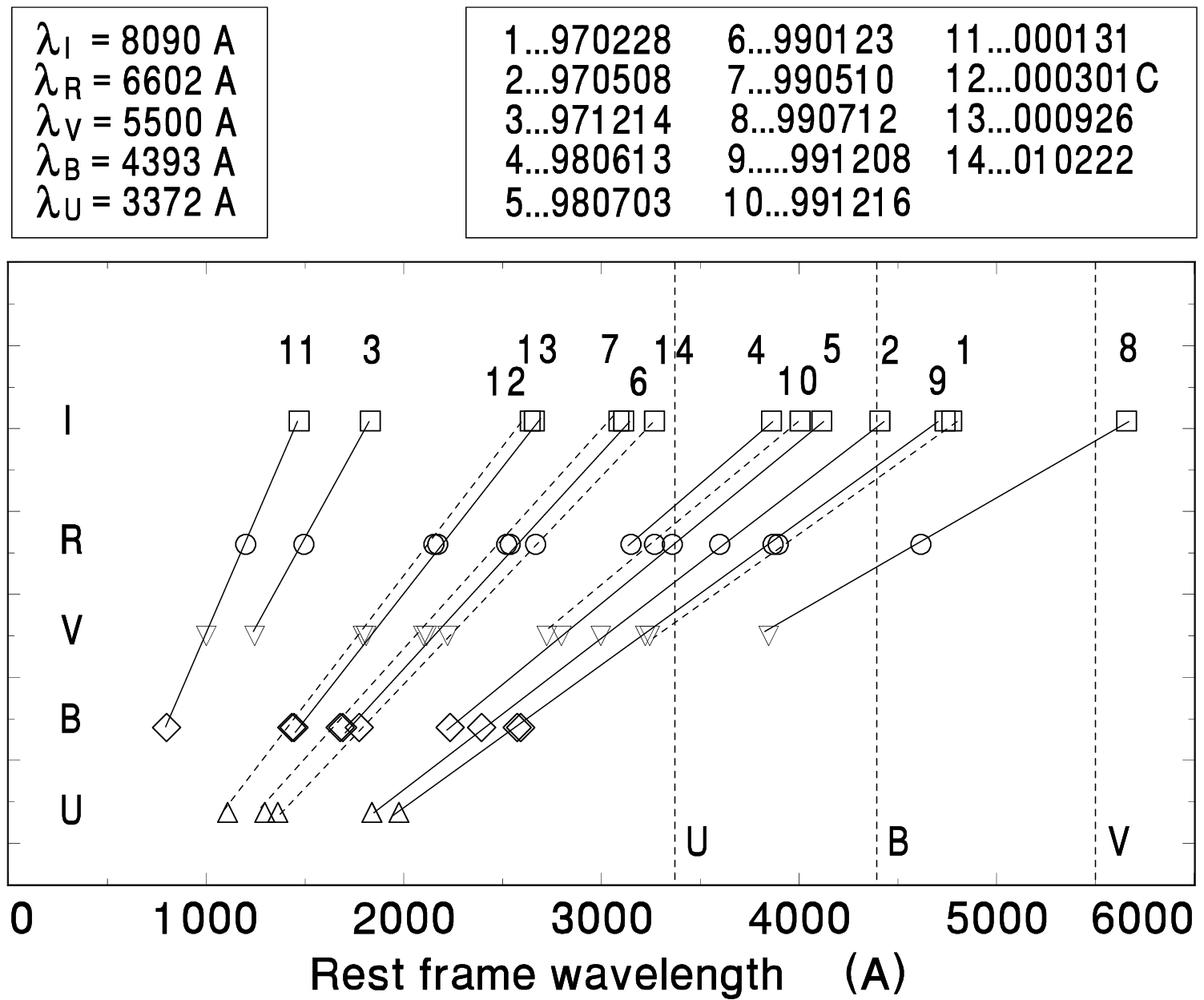,angle=0,width=9.2cm}
%\vspace{5cm}
\caption[ ]{Relation between the wavelength of radiation observed in a 
given  passband and the rest-frame wavelength, radiated by the OA with 
known redshift  $z$  ($U$-filter~-- triangles; $B$-filter~-- diamonds; 
$V$-filter~-- inverted triangles; $R$-filter~-- circles; $I$-filter~-- 
squares). Only those OAs for which at least one observation in a given 
passband exists  are  included  in  the plot. See Sect.\,\ref{red} for 
details.}
\label{lambd-aa}
\end{figure}

     In order to  determine meaningful color indices from the light curves of 
the OAs which  often  suffer from the shortcomings mentioned above, the light 
curves of  the  individual  OAs  in  all  available passbands were plotted in 
linear  time scale and critically examined. It emerged that in most cases the 
light curves are  free  of  complicated significant rapid changes on the time 
scale  of  hours to  a few days. The  mean  course  of  the light curve could 
therefore be  determined. If  several  measurements  in  a given  filter were 
obtained  within  a few  hours  and  had  comparable  standard  deviations of 
brightness  (quoted in the literature) then a centroid was calculated. If the 
standard  deviations  were  largely  discordant  then  just the most accurate 
measurements  were  selected  for further analysis. It emerged that the color 
indices could be determined directly from the measurements (centroids) if the 
observations  in  the respective  passbands  were  secured  within  about  an 
hour in  early  phase  of  the  OA ($t-T_{\rm 0}<3$ days)  and  even within a 
single  night  in  later  phases. In  some cases an interpolation between the 
measurements, obtained  in  the  neighbouring  nights, was used, particularly 
when the coverage of the light curve in some passbands was denser than in the 
others. The  standard  deviation  of each color index was calculated from the 
errors of  the  observations quoted in the literature. This procedure enabled 
us to  obtain  at  least  one  color  index  for  each  of the 17 afterglows. 
Examination of the light curves of the OAs in the respective filters revealed 
that the available data allow reliable curves to  be  constructed  in a given 
passband for $t-T_{\rm 0}<10$ days in most  cases (except  for  the  $R$-band 
which enabled to trace the evolution of  the  OAs to  fainter  brightness and 
hence to longer $t-T_{\rm 0}$). We therefore decided to limit our analysis of 
the  color indices to the points obtained  within  $t-T_{\rm 0}<10$ days. The 
color indices were corrected for the Galactic reddening according to the maps 
by Schlegel  et al. (1998). These color  indices  of the OAs along with their 
standard deviations are listed in Table~\ref{data}.

     The color indices  of the OAs were not corrected for the contribution of 
the host galaxies. This  can  be justified by the following reasons: we limit 
our analysis  of  the  colors  to the points obtained within $t-T_{\rm 0}<10$ 
days where the OAs were brighter than their hosts by several magnitudes, that 
means that the light contribution of the host galaxies is quite small and can 
be neglected.

     Also, we did not take into account  any redshift correction, which would 
be possible only if we  had  a correct  a priori  knowledge  of  the emerging 
spectrum of the OAs. Nevertheless, we  will  show that, despite the lack of a 
redshift  correction, the  colors  of most OAs cluster around well determined 
values.

\begin{table*}%[th]
\caption[ ]{The color indices of the optical afterglows of the GRBs,  
determined from the  available  data. Julian Date of the color index 
in the  form JD$-$2\,400\,000 is  given along with the time interval 
$t-T_{\rm 0}$ in  the observer frame, elapsed from the corresponding  
GRB  trigger. Only  the  color indices for $t-T_{\rm 0}<10$ days are 
listed. The standard  deviation  of  a given color index is given in 
the  line  below. The  color  indices are corrected for the Galactic 
reddening. The  last  two  columns  bring  the  redshift $z$ and the 
value of the interstellar Galactic reddening $E_{\rm B-V}$. }   
\label{data}
\[
\begin{array}{llcccccccll}
\hline
\noalign{\smallskip}
 $GRB$  &  $JD$    & t-T_{0} & (U-B)_{0} & (B-V)_{0} & (V-R)_{0} & (R-I)_{0} & (I-J)_{0} & (J-K)_{0} & z & E_{\rm B-V} \\
\hline
970228  &  50508.5  & 0.876   &          &  0.88     &  0.26     &  0.14     &           &        &  0.695  &  0.2167  \\
        &           &         &          &  \pm0.41  &  \pm0.18  &  \pm0.18  &           &        &         &          \\
\hline
970508  &  50578.42 & 1.018   &          &           &  0.15     &  0.35     &           &        &  0.835  &  0.0494  \\
        &           &         &          &           &  \pm0.07  &  \pm0.19  &           &        &         &          \\
        &  50579.27 & 1.868   &          &  0.39     &  0.33     &  0.48     &           &        &         &          \\
        &           &         &          &  \pm0.04  &  \pm0.04  &  \pm0.05  &           &        &         &          \\
        &  50579.47 & 2.068   &  -0.94   &  0.40     &  0.34     &  0.48     &           &        &         &          \\
        &           &         &  \pm0.26 &  \pm0.09  &  \pm0.07  &  \pm0.06  &           &        &         &          \\
        &  50580.25 & 2.848   &          &  0.46     &  0.39     &  0.49     &           &        &         &          \\
        &           &         &          &  \pm0.05  &  \pm0.04  &  \pm0.05  &           &        &         &          \\
        &  50580.52 & 3.118   &  -1.28   &  0.45     &  0.38     &  0.55     &           &        &         &          \\
        &           &         &  \pm0.26 &  \pm0.09  &  \pm0.09  &  \pm0.19  &           &        &         &          \\
        &  50581.35 & 3.948   &          &  0.33     &  0.44     &  0.41     &           &        &         &          \\
        &           &         &          &  \pm0.07  &  \pm0.06  &  \pm0.08  &           &        &         &          \\
        &  50582.40 & 4.998   &          &  0.40     &  0.35     &  0.48     &           &        &         &          \\
        &           &         &          &  \pm0.09  &  \pm0.09  &  \pm0.11  &           &        &         &          \\
\hline
971214  &  50798.50 & 1.025   &          &           &  0.59     &  0.56     &  0.95     &  1.45  &  3.418  &  0.016   \\
        &           &         &          &           &  \pm0.13  &  \pm0.07  &  \pm0.20  & \pm0.28 &        &          \\
        &  50799.50 & 2.025   &          &           &           &  0.45     &           &        &         &          \\
        &           &         &          &           &           &  \pm0.35  &           &        &         &          \\
        &  50800.50 & 3.025   &          &           &           &  0.68     &           &        &         &          \\
        &           &         &          &           &           &  \pm0.30  &           &        &         &          \\
\hline
980519  &  50953.93 & 0.93    &          &  0.48     &  0.51     &  0.56     &           &        &         &  0.2667  \\
        &           &         &          &  \pm0.24  &  \pm0.17  &  \pm0.12  &           &        &         &          \\
\hline
980613  &  50978.70 & 1.000   &          &           &           &  0.37     &           &        &  1.096  &  0.087   \\
        &           &         &          &           &           &  \pm0.13  &           &        &         &          \\
\hline
980703  &  51005.0  & 7.315   &          &  0.09     &  0.47     &  0.78     &           &  1.64  &  0.966  &  0.058   \\
        &           &         &          &  \pm0.14  &  \pm0.14  &  \pm0.14  &           & \pm0.18 &        &          \\
\hline
990123  &  51203.00 & 1.092   &  -0.87   &  0.32     &  0.31     &  0.44     &           &        &  1.60   &  0.016   \\
        &           &         &  \pm0.11 &  \pm0.07  &  \pm0.07  &  \pm0.09  &           &        &         &          \\
        &  51203.54 & 1.632   &          &  0.66     &  0.29     &           &           &        &         &          \\
        &           &         &          &  \pm0.26  &  \pm0.22  &           &           &        &         &          \\
\hline
990308  &  51245.85 & 0.140   &          &  0.33     &  0.18     &           &           &        &         &  0.023   \\
        &           &         &          &  \pm0.24  &  \pm0.09  &           &           &        &         &          \\
\hline
990510  &  51309.86 & 0.990   &          &  0.32     &  0.31     &  0.26     &           &        &  1.619  &  0.227   \\
        &           &         &          &  \pm0.06  &  \pm0.10  &  \pm0.10  &           &        &         &          \\
        &  51310.50 & 1.630   &          &           &  0.29     &  0.36     &           &        &         &          \\
        &           &         &          &           &  \pm0.10  &  \pm0.10  &           &        &         &          \\
\hline
990712  &  51374.5  & 2.300   &          &           &  0.68     &  0.37     &           &        &  0.430  &  0.0327  \\
        &           &         &          &           &  \pm0.25  &  \pm0.10  &           &        &         &          \\
\hline
991208  &  51523.75 & 3.060   &          &           &           &  0.44     &           &        &  0.707  &  0.016   \\
        &           &         &          &           &           &  \pm0.11  &           &        &         &          \\
        &  51524.77 & 4.080   &          &  0.59     &  0.44     &  0.36     &           &        &         &          \\
        &           &         &          &  \pm0.07  &  \pm0.07  &  \pm0.06  &           &        &         &          \\
        &  51525.78 & 5.090   &          &  0.66     &  0.51     &  0.56     &           &        &         &          \\
        &           &         &          &  \pm0.09  &  \pm0.08  &  \pm0.07  &           &        &         &          \\
        &  51526.77 & 6.080   &          &  0.63     &  0.27     &  0.50     &           &        &         &          \\
        &           &         &          &  \pm0.09  &  \pm0.08  &  \pm0.07  &           &        &         &          \\
        &  51527.77 & 7.080   &          &           &           &  0.51     &           &        &         &          \\
        &           &         &          &           &           &  \pm0.18  &           &        &         &          \\
\hline
991216  &  51529.72 & 0.55    &          &           &           &           &           & 1.46    &  1.02  &  0.52    \\
        &           &         &          &           &           &           &           & \pm0.10 &        &          \\
        &  51529.90 & 0.73    &          &           &           &           &           & 1.50    &        &          \\
        &           &         &          &           &           &           &           & \pm0.08 &        &          \\
\noalign{\smallskip}
\hline
\end{array}
\]
\end{table*}

\begin{table*}
{\bf Table~2}~-- continued
\[
\begin{array}{llcccccccll}
\hline
\noalign{\smallskip}
 $GRB$  &  $JD$    & t-T_{0} & (U-B)_{0} & (B-V)_{0} & (V-R)_{0} & (R-I)_{0} & (I-J)_{0} & (J-K)_{0} & z & E_{\rm B-V} \\
\hline
991216  &  51530.75 & 1.58    &          &           &           &           &           & 1.29    &        &          \\
        &           &         &          &           &           &           &           & \pm0.10 &        &          \\
        &  51530.87 & 1.70    &          &           &           &           &           & 1.46    &        &          \\
        &           &         &          &           &           &           &           & \pm0.11 &        &          \\
        &  51530.90 & 1.73    &          &           &  0.42     &  0.27     &  0.92     &         &        &          \\
        &           &         &          &           &  \pm0.08  &  \pm0.11  &  \pm0.12  &         &        &          \\
        &  51531.80 & 2.63    &          &           &           &           &           & 1.45    &        &          \\
        &           &         &          &           &           &           &           & \pm0.27 &        &          \\
\hline
000131  &  51578.64 & 3.525   &          &  >1.13    &  1.36     &  1.19     &           &        &  4.5    &  0.056   \\
        &           &         &          &           &  \pm0.11  &  \pm0.08  &           &        &         &          \\
\hline
000301C &  51606.70 & 1.788   &  -0.56   &  0.42     &  0.34     &  0.21     &           &        &  2.0404 &  0.053   \\
        &           &         &  \pm0.09 &  \pm0.06  &  \pm0.06  &  \pm0.08  &           &        &         &          \\
        &  51606.75 & 1.838   &          &           &           &  0.52     &           &        &         &          \\
        &           &         &          &           &           &  \pm0.08  &           &        &         &          \\
        &  51607.90 & 2.988   &  -0.57   &  0.25     &  0.43     &  0.50     &           &        &         &          \\
        &           &         &  \pm0.12 &  \pm0.09  &  \pm0.08  &  \pm0.11  &           &        &         &          \\
        &  51608.67 & 3.758   &          &  0.51     &  0.51     &  0.22     &           &        &         &          \\
        &           &         &          &  \pm0.28  &  \pm0.21  &  \pm0.14  &           &        &         &          \\
        &  51609.66 & 4.748   &          &           &           &  0.79     &           &        &         &          \\
        &           &         &          &           &           &  \pm0.25  &           &        &         &          \\
        &  51610.68 & 5.768   &  -0.55   &           &           &  0.40     &           &        &         &          \\
        &           &         &  \pm0.12 &           &           &  \pm0.21  &           &        &         &          \\
        &  51611.67 & 6.758   &  -0.47   &           &           &  0.22     &           &        &         &          \\
        &           &         &  \pm0.14 &           &           &  \pm0.10  &           &        &         &          \\
        &  51612.70 & 7.788   &  -0.36   &           &           &  0.37     &           &        &         &          \\
        &           &         &  \pm0.23 &           &           &  \pm0.21  &           &        &         &          \\
\hline
000911  &  51806.78 & 7.968   &          &           &  0.46     &  0.48     &           &        &         &  0.12    \\
        &           &         &          &           &  \pm0.09  &  \pm0.09  &           &        &         &          \\
\hline
000926  &  51815.69 & 1.191   &          &           &           &  0.51     &           &        &  2.066  &          \\
        &           &         &          &           &           &  \pm0.05  &           &        &         &          \\
        &  51816.24 & 1.741   &          &           &  0.73     &           &           &        &         &          \\
        &           &         &          &           &  \pm0.15  &           &           &        &         &          \\
        &  51816.70 & 2.201   &          &  0.62     &  0.40     &  0.86     &           &        &         &          \\
        &           &         &          &  \pm0.09  &  \pm0.07  &  \pm0.07  &           &        &         &          \\
        &  51817.70 & 3.201   &          &  0.70     &  0.51     &  0.83     &           &        &         &          \\
        &           &         &          &  \pm0.34  &  \pm0.34  &  \pm0.10  &           &        &         &          \\
\hline
010222  &  51963.69 & 0.890   &  -0.22   &  0.35     &  0.37     &  0.48     &   0.79    & 1.00    &  1.475 &  0.02    \\
        &           &         &  \pm0.09 &  \pm0.13  &  \pm0.10  &  \pm0.08  &   \pm0.26 & \pm0.39 &        &          \\
        &  51964.74 & 1.940   &  -0.32   &  0.41     &  0.40     &  0.59     &   1.26    & 1.72    &        &          \\
        &           &         &  \pm0.05 &  \pm0.04  &  \pm0.04  &  \pm0.05  &   \pm0.35 & \pm0.43 &        &          \\
        &  51965.76 & 2.960   &          &  0.51     &  0.30     &  0.49     &           &         &        &          \\
        &           &         &          &  \pm0.05  &  \pm0.05  &  \pm0.07  &           &         &        &          \\
        &  51967.76 & 4.960   &          &  0.48     &  0.42     &           &           &         &        &          \\
        &           &         &          &  \pm0.10  &  \pm0.08  &           &           &         &        &          \\
\noalign{\smallskip}
\hline
\end{array}
\]
\end{table*}

\subsection{The color diagrams of the afterglows  \label{color} }

      Fig.\,\ref{t-color}abcd shows  the  available  color indices $(R-I)_0$, 
$(V-R)_0$,  $(B-V)_0$  and  $(U-B)_0$  of  the  17 afterglows, plotted versus 
the  time interval  $t-T_{\rm  0}$  in  the observer frame, elapsed since the 
corresponding  GRB event  at $T_{\rm 0}$. $T_{\rm 0}$ refers to the moment of 
the onset of the GRB. Only  indices of  the  OAs  within 10 days from the GRB 
trigger  are  plotted along  with their standard deviations. $t-T_{\rm 0}$ is 
measured  in the  observer  frame  here  in  order  to include also the color 
indices of  those  OAs  for which no redshift $z$ is available. The colors of 
all OAs  correspond  to  their  typical  power-law  decline  branch, with the 
exception  of  the  first  point of GRB970508. The color indices of the whole 
ensemble of OAs, displayed in a given panel, were  also  merged into a common 
file and fitted  with a linear function to see if any evolution occurs in the 
first 10 days after the GRB. This fit is also shown in Fig.\,\ref{t-color}abc 
as the thick solid line along with its standard deviation (thin solid lines). 
The  data  in  Fig.\,\ref{t-color}d  were  not fitted  because  the available 
$(U-B)_0$  indices  of  the  OAs are not numerous enough to enable meaningful 
fitting. Because the  properties  of  SN1998bw,  the  possible counterpart of 
GRB980425, are  markedly  different  from  the  remaining OAs, its colors are 
shown  just for comparison  and were not included in  the fits. The colors of 
SN1998bw  were  determined from  the light  curves presented by Galama et al. 
(1998b). {\it The color  indices  of  all  OAs,  except  GRB000131,  occupy a 
narrow belt in Fig.\,\ref{t-color}abc  and the fits clearly  demonstrate that 
the  evolution  of  $(R-I)_0$,  $(V-R)_0$  and  $(B-V)_0$  of  these  OAs  is 
negligible over the considered  time interval}. The  slight non-zero slope of 
the fits is  of the order  of  0.2~mag  per  10~days  and stays  within their 
standard deviation. Although  the  color  indices  display some scatter, they 
mostly lie within the observational errors. We are therefore very cautious in 
drawing  strong conclusions about the "fine structure" of the color evolution 
of the OAs from the available data. Instead, we will concentrate  on analysis 
of the comprehensive properties of the  whole ensemble  of the color indices. 
The fact  that  the  color  indices  of  the  OAs  occupy  a  narrow  belt in 
Fig.\,\ref{t-color}abc suggests that common color-color diagrams can be built 
and  that  these  diagrams  are  meaningful  even  if the observations of the 
various  OAs  come from different epochs after the GRB, albeit within 10 days 
from it.

     Various  color-color  diagrams  for  the  OAs,  made with  the data from 
Table~\ref{data},  are  shown  in  Figs.\,\ref{vrri-aa},   \ref{bvvr-aa}  and 
\ref{bvub-aa}.  These  diagrams  have  identical  scales  of  the   axes (and 
Fig.\,\ref{vrri-aa} and~\ref{bvvr-aa} also the zero points) to allow a direct 
comparison of the scatter  among  the plots. Usually both color indices which 
form  a pair  in the color-color diagrams come from measurements separated by 
at most one  day. The  mean  colors  (centroid)  of the whole ensemble of OAs 
displayed in  each  diagram, including the standard deviations, are marked by 
crosses  in  Fig.\,\ref{vrri-aa}  and~\ref{bvvr-aa}. The  error  bars  of the 
individual  color  indices, which are plotted in Fig.\,\ref{t-color}abcd, are 
not  repeated  in Figs.\,\ref{vrri-aa},  \ref{bvvr-aa}  and~\ref{bvub-aa}  to 
avoid overcrowding of the plots.

     Figs.\,\ref{vrri-aa},  \ref{bvvr-aa}  and  \ref{bvub-aa}  also  show the 
representative  reddening  paths  for  $E_{\rm B-V}=0.5$.  The  values of $z$ 
differ for  the  respective  OAs such that observations in a given filter can 
therefore comprise radiation within a large range  of wavelengths in the rest 
frame  (see  Sect.\,\ref{color}). Because  the  observer  in a given passband 
will detect radiation  at  progressively shorter  wavelengths with increasing 
redshift  $z$, we  decided  to  include  the reddening paths, appropriate for 
$U-B$, $B-V$, $V-R$ and $R-I$, in all color-color diagrams. It can clearly be 
seen  that both the lengths and directions of the vectors are similar for all 
reddening paths.

\vspace{2ex}

  $V-R$ vs. $R-I$  diagram  (Fig.\,\ref{vrri-aa}): The  color  indices of all 
OAs, except  for  GRB000131, occupy  a well  localized region of  the diagram 
and  display  no apparent  scatter  along  the reddening path. Colors of some 
main-sequence  stars  which  are  also  included  can  be  used  for the {\it 
observational} differentiation of the OAs from other objects.

\vspace{1ex}

    $B-V$ vs. $V-R$ diagram (Fig.\,\ref{bvvr-aa}): All afterglows, except for 
GRB000131, occupy  just  a small  region  of  the  diagram again, without any 
scatter along the reddening path.

\vspace{1ex}

    $U-B$ vs.  $B-V$  diagram  (Fig.\,\ref{bvub-aa}):  Only  four  afterglows 
(GRB970508, GRB990123, GRB991208 and GRB000301C) allow  determining the $U-B$ 
index or at least  its limit. Notice the much larger scatter of the events in 
the  $U-B$  direction  than  in $B-V$. Positions  of the main sequence stars, 
giants  and  the locus of black bodies are plotted~-- they are useful for the 
{\it observational} differentiation of OAs from other objects.

\vspace{1ex}

    $I-J$ and  $J-K$ indices: They are available only for three and four OAs, 
respectively (Table~\ref{data}). It  can  be  seen from Table~\ref{data} that 
there is good agreement in $I-J$ and $J-K$ indices among all events for which 
these indices were measured.

\vspace{1ex}

           Figs.\,\ref{vrri-aa}  and  \ref{bvvr-aa}  then  indicate a similar 
distribution for the  optical  colors of this OA sample ($B-V$=0.47$\pm$0.17, 
$V-R$=0.40$\pm$0.13  and  $R-I$=0.46$\pm$0.18).  These   average  colors  are 
consistent with a power-law shaped optical spectral distribution, in the form 
$F_\nu\propto\nu^{-\beta}$, with  $\beta\sim$1. This spectral shape naturally 
follows  from  the  theoretical  treatment  of  the  GRB  afterglow  emission 
suggested by Sari et al. (1998) in the framework of the `fireball' model (see 
Sect.\,\ref{dis} for details).

\subsection{Dependence of the color index on the redshift~$z$  \label{red} }

     The observed passbands  of OAs  will be different from those in the rest 
frame because  of  the  effects introduced by the redshift $z$. The values of 
$z$ of the OAs considered here lie in the range 0.43--4.5, typical  $z$ being 
around 1. In  the  case of  a complicated  spectrum  shape  the color indices 
would be a function  of  $z$. The color indices of all OAs with known $z$ are 
plotted  versus  $z$ in Fig.\,\ref{color-z2}.Only the  OAs  with  $z<3.5$ are 
considered  here; the case of GRB000131 ($z=4.5$) is discussed in more detail 
in Sect.\,\ref{dis}. The fits show that any dependence of the color on $z$ is 
weak and within the 1\,$\sigma$ errors for $(R-I)_0$, $(V-R)_0$ and $(B-V)_0$. 
The only exception may be $(U-B)_0$.

     The relation between the wavelength observed in a given passband and the 
rest-frame wavelength, radiated by the OA with known redshift $z$ can be seen 
in Fig.\,\ref{lambd-aa}. This  diagram  shows  that the $(R-I)_0$, $(V-R)_0$, 
$(B-V)_0$  and  $(U-B)_0$  indices  represent   the  radiation  between about 
800~\AA~and 5600~\AA~in the rest frame.

     The relation in  Fig.\,\ref{lambd-aa}, when  compared  with  the  strong 
clustering of the  color  indices  in Figs.\,\ref{vrri-aa} and~\ref{bvvr-aa}, 
clearly shows that  the  $(R-I)_0$, $(V-R)_0$ and $(B-V)_0$ indices represent 
the radiation between about 1500 and 5500~\AA~in the rest frame. On the other 
hand, the large scatter, observed  in $(U-B)_0$, originates below 1500~\AA~in 
the rest frame.

\subsection{Absolute brightness of the OAs  \label{abs} }

\begin{figure}[h!]
%\rule{0.4pt}{21.2cm}
%\rule{0.4pt}{1cm}
\hspace{-2.5cm}
\epsfig{figure=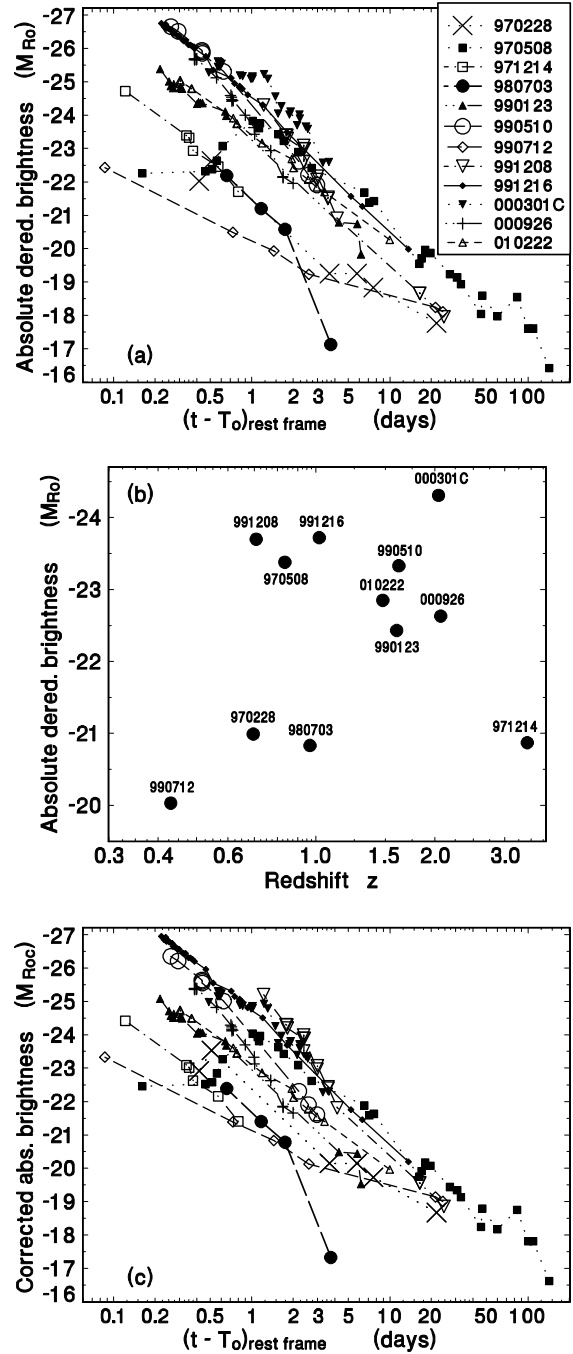,angle=-90,width=14cm}
%\vspace{-.5cm}
\caption[ ]{{\bf (a)}~Absolute  brightness  of  the  OAs  in the observed 
$R$-band, corrected for the Galactic reddening and the light contribution 
of the host  galaxy. The  time  intervals $t-T_{\rm 0}$  in  the observer 
frame were  transformed  into  the rest frame $(t-T_{\rm 0})_{\rm rest}$. 
{\bf (b)}~Absolute  brightness  $M_{\rm  R_0}$  of  the OAs,  measured at 
$(t-T_{\rm  0})_{\rm  rest}=1.5$~days,  plotted  versus  redshift  $z$. A 
correlation between $z$ and $M_{\rm R_0}$ would be expected if the spread 
of brightness  of  the  respective  light  curves  in Fig.\,a were caused 
purely  by  the  shift of $\lambda$. {\bf (c)}~The same as in Fig.\,a but 
zeroth order $k$-correction was applied to the observed $R$ magnitudes of 
the respective OAs. See Sect.\,\ref{abs} for details.}
\label{abs-mag3}
\end{figure}

     Absolute brightness in  the  $R$-band  ($M_{\rm R_0}$)  for  each OA was 
calculated in  two  steps:  first, by  determining  the  luminosity  distance 
$D_{\rm L}$  from  the  redshift $z$, using Eq.15.3.25 of Weinberg (1972) and 
$H_{\rm  0}~=  60$~km~s$^{\rm  -1}$  Mpc$^{\rm  -1}$;  then, by  applying the 
distance  modulus  formula  using  $D_{\rm  L}$  and  the  observed  $R$-band 
magnitude  corrected  for  the  Galactic  reddening  according to the maps by 
Schlegel  et  al. (1998). The  time intervals $t-T_{\rm 0}$ in  the  observer 
frame  were transformed into the rest frame times $(t-T_{\rm 0})_{\rm rest}$. 
The  resultant  light  curves  for  OAs  with  known  $z$  are  displayed  in 
Fig.\,\ref{abs-mag3}a. Here the  brightness  of  OAs in Fig.\,\ref{abs-mag3}a 
was  also  corrected  for  the  light contribution of the host galaxy because 
these  $R$-band  light  curves extend  far behind $t-T_{\rm 0}<10$ days, used 
as a limit  for the  color indices, and the brightness of the OA decreased so 
much that  the  light  of  the host could not be neglected. We note, however, 
that in  most  cases  the host galaxy noticeably affects the brightness of OA 
only in its very late phase ($t-T_{\rm 0}$ of the order of tens of days).

     It  is  desirable  to  examine   whether  the  spread  of  the  absolute 
brightnesses $M_{\rm  R_0}$ of the respective OAs in Fig.\,\ref{abs-mag3}a is 
real or if it  is  caused  mainly  by  the shifts of the effective wavelength 
$\lambda$ due to the  different  redshifts  $z$. The observer in the $R$ band 
will detect radiation originated at  progressively shorter wavelengths in the 
GRB rest frame with increasing redshift $z$. As it  will be argued below, the 
spectra  of  the  OAs are very smooth and very similar each to other. Because 
generally the intensity may be somewhat different at various $\lambda$ due to 
the  slope  of  the  spectrum, the  shift  of  the  effective  $\lambda$ will 
introduce  some  change  of  brightness. K-correction is possible only if the 
shape  of the  spectrum is  known  a priori. First, we decided to make a {\it 
model-independent}  test which does not impose  any assumptions  on the shape 
and  the slope of the spectrum of OAs. A correlation between the redshift $z$ 
and $M_{\rm  R_0}$  (measured for some $(t-T_{\rm 0})_{\rm  rest}$, identical 
for all OAs)  is  expected  if the spread of brightness were caused purely by 
the  shift  of  $\lambda$. $M_{\rm R_0}$  was measured at the same $(t-T_{\rm 
0})_{\rm  rest}=1.5$~days. All OAs, plotted in Fig.\,\ref{abs-mag3}a, already 
achieved the  decay  branch  of  their  light  curve. The light curves of the 
decays  of  all OAs in Fig.\,\ref{abs-mag3}a were fitted by the straight line 
because only the mean slope, averaged over the "fine" structure  seen in some 
events, was  investigated. $M_{\rm R_0}$  was  then read out from the fit. An 
extrapolation of the decay had to be used for the OA of GRB971214. The result 
is shown in  Fig.\,\ref{abs-mag3}b. It  can readily be  seen that there is no 
apparent correlation between $z$ and $M_{\rm R_0}$.

     A further correction of the $M_{R_0}$ of the OAs in this sample has been 
performed  by  assuming  that  the OA  spectra  can  be modeled using a power 
law with  index  $\beta\sim$1,  as  suggested  by  the  results  reported  in 
Sect.\,\ref{color},  and  keeping  into  account  which  rest-frame   band is 
corresponding  to  the $R$ band as seen in the observer frame. In particular, 
as the  observer  frame $R$ bands of the OAs in this sample mainly lie within 
the $B$ optical band  and  around  2000~\AA~in the UV region (e.g. the $UVW2$ 
filter  of  the  Optical  Monitor onboard  the  XMM-Newton  satellite; Dahlem 
2000), we  first computed  the  color indices $UVW2-R$, $U-R$, $B-R$ assuming 
$\beta\sim$1. In  this  hypothesis,  they  are  $\sim$$-$0.3  (the rest-frame 
wavelength  of  the  OAs  $\lambda_{\rm  rest}~=  1400-2800$~\AA),  $\sim$0.2 
($\lambda_{\rm rest}~=  2800-3750$~\AA) and  $\sim$0.9 ($\lambda_{\rm rest}~= 
3750-4800$~\AA),  respectively. Next, we  corrected the observed absolute $R$ 
magnitude  by  subtracting  the  appropriate color index value from $M_{R_0}$ 
as determined by  using  the  luminosity  distance  and  the distance modulus 
formulae. This is some sort of zeroth order $k$-correction introduced to make 
an attempt  at  taking  into  account  the  shift  on  the OA optical spectra 
produced  by the cosmological redshift. In this way we can better compare the 
`real' $R$-band absolute magnitudes of this set of OAs (Fig.\,\ref{abs-mag3}c).

\section{Discussion    \label{dis}    }

     We have shown that the color indices of most OAs occupy the well defined 
belts, when  plotted  as  a function of time, especially for $t-T_{\rm 0}<10$ 
days, and  that the time evolution of these indices is negligible during this 
interval. Most afterglows appear to concentrate in the well localized regions 
of $V-R$ vs. $R-I$ and $B-V$ vs. $V-R$ diagrams. Colors of  all  OAs analyzed 
here correspond to their final decline branch with the exception of the first  
point  of GRB970508. The color  indices  $(R-I)_0~=  0.46\pm0.18$, $(V-R)_0~= 
0.40\pm0.13$, $(B-V)_0~= 0.47\pm0.17$  appear  to  be  typical. OAs therefore 
appear  quite  red  in  the  spectral region  between the  $I$ and $B$-filter 
(colors  similar to mid-G or early K stars). On the other hand, the $(U-B)_0$ 
index  of the OAs is negative with a large  scatter  and  the position of the 
afterglows  in $U-B$ vs. $B-V$ diagram  is often  quite  different from other 
sources. These specific colors can be used to distinguish genuine OAs related 
to GRBs from another types of objects in optical GRB searches.

     The results reported here  are thus extremely significant from the point 
of view  of  optical  searches  for  $\gamma$-ray  bursters.  First, the well 
defined color  of  OAs  allows,  together  with  the  power  law  decline, to 
distinguish  the  real OAs  of  GRBs  from  the other  types of astrophysical 
objects. So far, only the rapid fading behavior has served to distinguish the 
genuine  afterglows. So  we  have  another  tool now, enabling one to analyze 
images  taken  in  various  optical  passbands  quickly  for possible optical 
counterparts  of  GRBs, without  waiting  for  the following night to confirm 
the  rapid  object  fading. Secondly,  the  well  focused  position of OAs in 
color-color diagrams represents a very important tool to consider independent 
(on satellite projects) optical searches for OAs of  GRBs. Although there are 
suitable databases available for this kind of analyses such as the deep UKSTU 
archival  plates, the  search  for real OAs  was  difficult  because  of high 
background level of other types of variable phenomena  not  related  to GRBs. 
The combination of fading profile with the color information may be used as a 
very powerful tool to search for real OAs in  suitable  databases  with color 
information  such  as  the  UKSTU  archive.  It  should  be  noted  that  the 
actual  rate  of  OAs may  exceeds  those of GRBs due to different beaming in 
$\gamma$-rays  and  in optical  (Hudec  2000). Moreover, the  OAs detected by 
optical  searches  may  provide very precise localizations of GRBs, and hence 
allow detailed  studies  of  their  host galaxies. On the other hand, the OAs 
rates and/or limits provided by these analyses could  provide  constraints on 
the time and/or wavelength dependent beaming in GRBs, and hence contribute to 
the understanding of the physical model of the events.

     The  results  presented  here  also  confirm  the  importance  of  color 
information  in  the strategy of the optical afterglow searches and analyses. 
Not  only  the  fading  profile, but  also  the color information may yield a 
valuable  physical  conclusion  regarding  the  model, the  origin,  and  the 
position of  GRBs. We  hence  stress  that it is important to provide further 
optical  observations of  OAs  in  a well  organized  way. If  possible,  the 
observations  should  be carried  out  with  standard  filters,  and  various 
passbands should be taken immediately i.e. during the same night.

\vspace{1ex}

     The fact that most OAs  concentrate  in  the well defined regions of the 
color-color  diagrams  with the standard deviations just about 0.13--0.18~mag 
(except  for  $(U-B)_0$), despite  the  smearing  introduced by the different 
redshifts $z$ and the corresponding shifts of the passbands, implies that the 
spectral shape  of  OAs  is  very  smooth,  with  no  bumps  or strong lines, 
within the observed $I$ to $B$ passbands. The relation between the wavelength 
of  radiation  observed  in  a given  passband and the rest-frame wavelength, 
radiated  by  the  OA  (Fig.\,\ref{lambd-aa}),  shows   that  the  $(R-I)_0$, 
$(V-R)_0$, $(B-V)_0$  indices which display very small scatter, represent the 
radiation within about 2000 and 5600~\AA~in the rest frame. Also the slope of 
the  spectrum  can  be supposed to be almost constant and may have a powerlaw 
shape  $F_\nu\propto\nu^{-\beta}$  in  this  spectral  region. If  $\beta$ is 
similar  for  practically  all  OAs, then  the  redshift  effect on the ratio 
between luminosities  in  different  wavebands (i.e. color  indices)  is very 
small, as observed. Indeed, the  average  colors  in Figs.\,\ref{vrri-aa} and 
\ref{bvvr-aa}  are  consistent  with  a  power-law  shaped  optical  spectral 
distribution with  $\beta\sim$1. This  spectral  shape naturally follows from 
the  theoretical treatment of the GRB afterglow emission suggested by Sari et 
al. (1998)  in  the  framework  of  the `fireball' model. In  this  model the 
luminosity comes  from  a relativistic expanding and decelerating shell which 
radiates  via  synchrotron  emission. The  value  found  for $\beta$ is fully 
compatible, within this model, with an electron power-law energy distribution 
with  index  $p$$\sim$2.5, as  observationally  found by e.g. Frontera et al. 
(2000) from the spectral analysis of GRB prompt high-energy emission.

     The available  color  indices of the OAs represent relatively late stage 
of the event  because  they  come  from  $t-T_{\rm 0}\geq0.14$~days. They are 
thus expected to represent the blast  wave already moving rather spherically, 
with just  a little  beaming  (Piran 1999). In  the framework of the fireball 
model, these color indices represent  the processes in which only the forward 
external  shocks  play a role. The  similarity  of  the color indices of most 
OAs considered here suggests that the properties of these shocks remain quite 
similar for the respective events.

     It  is  remarkable  that  the  color  indices  $(R-I)_0$,  $(V-R)_0$ and 
$(B-V)_0$  do  not  evolve within $t-T_{\rm 0}<10$ days (Fig.\,\ref{t-color}) 
although the brightness of all OAs declines by several magnitudes during this 
time  interval  (Fig.\,\ref{abs-mag3}ac). It  implies  that  the shape of the 
spectrum does  not  change  significantly  while  the  luminosity  of the OAs 
decreases by a large amount.

     The absence  of  correlation between $z$ and $M_{\rm R_0}$ together with 
the very small scatter  of the color indices of the respective OAs allows one 
to conclude that the smearing of  $M_{\rm  R_0}$, introduced by the different 
values of $z$, does  not  significantly alter  the  scatter of  $M_{\rm R_0}$ 
of  the   various   events.  The  range  of  $M_{\rm  R_0}$  of  the  OAs  in 
Fig.\,\ref{abs-mag3}c for $(t-T_{\rm 0})_{\rm rest}=0.25$ days where most OAs 
already  reached  their  final decline branch is about 4.3 mag, from  $M_{\rm 
R_0}=-26.5$ to $-22.2$. It implies the range of luminosities about 1:50. 

     The fact that the  spectra  are  similar  although the luminosity of OAs 
appears to be different is most likely due to the total initial energy of the 
afterglow. As a matter of  fact, the spectral shape of the fireball model, as 
modeled  by Sari et al. (1998), does not depend on the input energy while the 
luminosity  of the afterglow at a particular epoch does depend on it. So, the 
higher  the  GRB  input  energy  is, the  brighter the OA is; this would also 
suggest one more time  that  GRBs and their afterglows {\it are not} standard 
candles.

     In any case, although  it is  apparent also from  Fig.\,\ref{abs-mag3}ac 
that the OAs  are  not standard candles, it  is interesting to note that five 
OAs have very similar $M_{\rm R_0}$  for $1<(t-T_{\rm  0})_{\rm rest}<3$~days  
(GRB970508, 990123, 990510, 991208, 000301C).

     The general  slope of the decay branches of OAs is largely independent 
of $M_{\rm R_0}$ and the  light  curves  of  all  OAs  considered  here are 
almost  parallel. This means that the difference in the luminosities of the 
respective  OAs  persists  through the decline, at least for $0.2<(t-T_{\rm 
0})_{\rm rest}<20$ days. The OA of GRB990712 might have a slower decay than 
the  others  but its host galaxy was exceptionally bright~-- the brightness 
of an  OA  is then  largely dependent on the exact  value of the brightness 
of the  host. The  different  observed  mean  decay slopes therefore can be 
attributed  mostly  to  the  relativistic effects~-- most these differences 
disappear when transformation to the rest frame is made.

\vspace{1ex}

     The strong  concentration  of  the  color  indices  in  the  color-color 
diagrams (Figs.\,\ref{vrri-aa} and~\ref{bvvr-aa}) suggests that the intrinsic 
reddening (i.e. in  their  host  galaxies)  must be quite similar for all OAs 
and,  moreover, that  this reddening is likely to be rather small. The reason 
is that in the case of a large reddening it would be quite unlikely to obtain 
such  similar  values  of  absorption in all cases. Notice that there is {\bf 
no} apparent scatter of the  color  indices of the OAs (except for GRB000131) 
along the reddening path, depicted in Figs.\,\ref{vrri-aa} and~\ref{bvvr-aa}. 
The scatter in the $U-B$ vs. $B-V$ diagram (Fig.\,\ref{bvub-aa}) is large but 
only  in  the  $U-B$ direction  which  is  inconsistent with the interstellar 
reddening. All  these  lines of evidence therefore imply that most GRBs whose 
afterglows are analyzed  here  are  unlikely  to come directly from the inner 
(densest) parts of the star-forming regions. However, that does  not  exclude 
the  possibility  that  these  GRBs  originate  on "our side" of a structured 
star-forming region. Alternatively, the density and the dust abundance of the 
local  interstellar  medium  might  be  substantially  reduced by the intense 
high-energy radiation  of  the  GRB  trigger, as  modeled by Waxman \& Draine 
(2000).

%and by Fruchter et al. (2001).

     We are however  aware  that in some cases the OAs appear to have steeper 
optical  spectra  (and  thus  redder  colors),  as,  for  example,  GRB980329 
(Palazzi  et  al.  1998),  GRB990705   (Masetti  et   al.  2000b),  GRB000131 
(Andersen  et  al. 2000), GRB000418 (Klose  et al. 2000), GRB000630 (Fynbo et 
al. 2000). These GRBs show $\beta\sim$2, or higher. This is  most  likely due 
to  strong  local  absorption  in  the  burst environment (as in the cases of 
GRB990705  and GRB000418)  or very  high  redshift  (as  for  GRB000131),  or 
possibly  both  (the  case  of  GRB980329;  Fruchter 1999). Thus, our results 
suggest  that the sample of OAs considered here is  very  little  affected by 
reddening effects induced by both strong local  absorption  or high redshift, 
i.e. here  we  are  dealing  with  GRB  afterglows which are at $z<4$ and not 
deep  inside  dense  dust  clouds in  their  host  galaxies. At  present, the 
available  data enable the determination of the color indices  of  the redder 
OAs  according  to  the  criteria  from Sect.\,\ref{col}  only  for the OA of 
GRB000131. Nevertheless, the strong concentration of the color indices of the 
OAs in  Figs.\,\ref{vrri-aa} and \ref{bvvr-aa} allows one to infer that there 
may not be  smooth  transition between the events considered in our  analysis 
and these OAs with steep optical spectra.

     The supernova  SN1998bw,  which  is  a possible  optical  counterpart of 
GRB980425, is markedly  different from the remaining OAs in several respects. 
While  the  OAs  from  Table~\ref{lit}  lie  at  the  cosmological  distances 
($z\geq0.43$), the  much smaller redshift of SN1998bw ($z=0.0085$) suggests a 
considerably  smaller distance and hence much lower absolute brightness. Also 
the light curve of SN1998bw largely  differs  from  the remaining OAs because 
the brightness of SN1998bw  was   still  steadily increasing within $t-T_{\rm 
0}<10$ days  while  all  other OAs already achieved their decline  branch and 
faded  by  several magnitudes. Generally, the  color  indices  of SN1998bw in 
Fig.\,\ref{t-color},  \ref{vrri-aa},  \ref{bvvr-aa} and \ref{bvub-aa} suggest 
the shape  of  spectrum  which  is different from the other OAs and cannot be 
explained purely by the different values of $z$.

     Comparison  of  the colors of OAs and their evolution with the behaviour 
of supernovae in general  may  be  fruitful  because  both  cases  presumably 
represent explosive  processes. At this stage, comparison of the {\it general 
trends} in  the  color  evolution  of  the  kinds of objects seems to be more 
suitable than a detailed  comparison  of  what  happens  on the absolute time 
scale. Let  us  confine  to  $t-T_{\rm 0}<10$~days for OAs (brightness of SNe 
typically falls by about the same amount after more than 100 days). The small 
spread  of  colors of OAs is a feature similar to the type Ia SNe where after 
a rapid  rise  (up  to  approx. 30  days  after  $T_{\rm  max}$) $B-V$ of the 
individual  SNIa  agrees  within  about 0.2~mag and slowly decreases linearly 
with time during a large part  of the  decline  ($30<t-T_{\rm  max}<90$ days) 
(Phillips et al. 1999). On the contrary, SNII display a large spread in $B-V$ 
and this index generally increases  by almost 1~mag during the first 100 days 
(Patat et al. 1994).

\begin{acknowledgements}
This  research  has made use of  NASA's  Astrophysics Data System Abstract 
Service. We acknowledge the support by the project KONTAKT ME~137 and ES02 
by the Ministry of Education and Youth of the Czech Republic and the grant 
205/99/0145 of the Grant Agency of the Czech Republic. We also acknowledge 
the CNR-AV\v{C}R Joint Research Program No.\,3 (1998/2000).
\end{acknowledgements}


\begin{thebibliography}{}

\bibitem{} Andersen, M., Hjorth, J., Pedersen, H., et al., 2000, A\&A, 
           364, L54

\bibitem{} Bloom, J.S., Frail, D.A., Kulkarni, S.R., et al., 1998,
           ApJ, 508, L21

\bibitem{} Castander, F.J., Lamb, D.Q., 1999, ApJ, 523, 593

\bibitem{} Castro-Tirado, A.J., Sokolov, V.V., Gorosabel, J., et al., 2001, 
           A\&A, 370, 398

\bibitem{} Dahlem M., 2000, XMM User's Handbook, version 1.2, available at
           {\tt http://xmm.vilspa.esa.es/user/uhb/XMM\_UHB}

\bibitem{} Diercks, A.H., Deutsch, E.W., Castander, F.J., et al., 1998,
           ApJ, 503, L105

\bibitem{} Frontera, F., Amati, L., Costa, E., et al., 2000, ApJS, 127, 59

\bibitem{} Fruchter, A.S., 1999, ApJ, 512, L1

\bibitem{} Fruchter, A.S., Pian, E., Thorsett, S.E., et al., 1999, ApJ,
           516, 683

\bibitem{} Fynbo J.U., Jensen, B.L., Gorosabel, J., et al., 2000, A\&A, 
	     369, 373

\bibitem{} Gorosabel, J., Castro-Tirado, A.J., Willott, C.J., et al., 1998, 
           A\&A, 335, L5 

\bibitem{} Galama, T.J., Groot, P.J., van Paradijs, J., et al., 1998a,
           ApJ, 497, L13

\bibitem{} Galama, T.J., Vreeswijk, P.M., van Paradijs, J., et al., 1998b,
           Nat, 395, 670

\bibitem{} Galama, T.J., Briggs, M.S., Wijers, R.A.M.J., et al., 1999,
           Nat, 398, 394

\bibitem{} Guarnieri, A., Bartolini, C., Masetti, N., et al., 1997,
           A\&A, 328, L13

\bibitem{} Halpern, J.P., Thorstensen, J.R., Helfand, D.J., Costa, E.,
           1998, Nat, 393, 41

\bibitem{} Halpern, J.P., Kemp, J., Piran, T., Bershady, M.A., 1999,
           ApJ, 517, L105

\bibitem{} Halpern, J.P., Uglesich, R., Mirabal, N., et al., 2000, ApJ,
           543, 697

\bibitem{} Hudec, R., 2000. In: Kippen R.M., Mallozzi R.S., Fishman G.J. 
           (eds.) Gamma-Ray Bursts - 5th Huntsville Symposium, AIP 
           Conf. Proc. 526,

\bibitem{} Jensen, B.L., Fynbo, J.U., Gorosabel, J., et al., 2000, A\&A,
           370, 909

\bibitem{} Klose, S., Stecklum, B., Masetti, N. et al., 2000, ApJ, 545, 271

\bibitem{} Kulkarni, S.R., Djorgoski, S.G., Ramaprakash, A.N., et al.,
           1998, Nat, 393, 35

\bibitem{} Masetti, N., Bartolini, C., Guarnieri, A., Piccioni, A., 1998.
           In: Scarsi L., Bradt H., Giommi P., Fiore F. (eds.) The Active 
           X-ray Sky: Results from BeppoSAX and Rossi-XTE, Nuc. Phys. B 
           Proc. Suppl. 69/1-3, 674 

\bibitem{} Masetti, N., Bartolini, C., Bernabei, S., et al., 2000a,
           A\&A, 359, L23

\bibitem{} Masetti, N., Palazzi, E., Pian, E., et al., 2000b, A\&A, 354, 473

\bibitem{} Masetti, N., Palazzi, E., Pian, E., et al., 2001, in press
	     (astro-ph/0103296)

\bibitem{} Palazzi, E., Pian, E., Masetti, N., et al., 1998, A\&A, 336, L95

\bibitem{} Patat, F., Barbon, R., Cappellaro, E., Turatto, M., 1994,
           A\&A, 282, 731

\bibitem{} Pedichini, F., Di Paola, A., Stella, L., et al., 1997,
           A\&A, 327, L36

\bibitem{} Phillips, M.M., Lira, P., Suntzeff, N.B., et al., 1999, AJ,
           118, 1766

\bibitem{} Piran, T., 1999, Phys. Rep., 314, 575

\bibitem{} Price, P.A., Harrison, F.A., Galama, T.J., 2001, ApJ, 549, L7

\bibitem{} Ramaprakash, A.N., Kulkarni, S.R., Frail, D.A., et al., 1998, 
           Nat, 393, 43 

\bibitem{} Rhoads, J.E., 2001, ApJ, 557, 943

\bibitem{} Sari, R., Piran, T., Narayan, R., 1998, ApJ, 497, L17

\bibitem{} Schaefer, B.E, Snyder, J.A., Hernandez, J., 1999, ApJ, 524, L103

\bibitem{} Schlegel, D.J., Finkbeiner, D.P., Davis, M., 1998, ApJ, 500, 525

\bibitem{} \v{S}imon, V., Pizzichini, G., Hudec, R., 2000a. In: Kippen R.M., 
           Mallozzi R.S., Fishman G.J. (eds.) Gamma-Ray Bursts - 5th Huntsville 
           Symposium, AIP Conf. Proc. 526,

\bibitem{} \v{S}imon, V., Hudec, R., Masetti, N., Pizzichini, G., 2000b, The 
           Procedings of the 4th INTEGRAL Workshop, 4--8 September, Alicante, 
           Spain (accepted)

\bibitem{} \v{S}imon, V., Hudec, R., Masetti, N., Pizzichini, G.,  2000c. In:
           Frontera, F., Costa, E., \& Hjorth, J., (eds.) Gamma-Ray Bursts 
           in the Afterglow Era: 2nd Workshop (A Joint CNR/ESO Meeting), 
           C.N.R. Headquarters, Rome, Italy (October 17-20, 2000)

\bibitem{} Sokolov, V.V., Kopylov, A.I., Zharikov, S.V., et al., 1998,
           A\&A, 334, 117

\bibitem{} van Paradijs, J., Groot, P.J., Galama, T.J., et al., 1997, Nat, 386, 
           686

\bibitem{} Vrba, F.J., Henden, A.A., Canzian, B., et al., 2000, ApJ, 528, 254

\bibitem{} Waxman, E., \& Draine, B.T., 2000, ApJ, 537, 796

\bibitem{} Weinberg, S., 1972, Gravitation and cosmology: principles and
           applications of the general theory of relativity. Wilby,
           New York, p. 485

\bibitem{} Zharikov, S.V., Sokolov, V.V., Baryshev, Yu.V., 1998, A\&A,
           337, 356


\end{thebibliography}
\end{document}